\documentclass[]{aastex}

\usepackage{emulateapj5}
\usepackage{onecolfloat}
\usepackage{graphicx}
\usepackage{fancyheadings}
\usepackage{ulem}
\usepackage{rotating}
\usepackage{lscape}

\newcommand{\msol}{M_\odot}

\newcommand{\lya}{Ly$\alpha$}

\newcommand{\ngrb}{14}
\newcommand{\nmg}{14}
\newcommand{\nqmg}{3.8}

\newcommand{\pth}{15.5}
\newcommand{\kms}{km~s$^{-1}$ }

\newcommand{\mkms}{{\rm \; km\;s^{-1}}}

\def\lick{1}
\def\uc{2}
\def\berk{3}
\def\geneve{4}
\def\chile{5}
\def\ioa{6}
\def\harvard{7}

\begin{document}

\twocolumn[%
\submitted{Submitted to ApJL}

\title{On the Incidence of Strong \ion{Mg}{2} Absorbers
Along GRB Sightlines }

\author{ 
	Gabriel E. Prochter\altaffilmark{\lick},
	Jason X. Prochaska\altaffilmark{\lick},
        Hsiao-Wen Chen\altaffilmark{\uc}, 
        Joshua S. Bloom\altaffilmark{\berk},
	Miroslava Dessauges-Zavadsky\altaffilmark{\geneve}, 
	Ryan J. Foley\altaffilmark{\berk}, 
	Sebastian Lopez\altaffilmark{\chile}, 
	Max Pettini\altaffilmark{\ioa}, 
        Andrea K. Dupree\altaffilmark{\harvard},
        P. Guhathakurta\altaffilmark{\lick}
        }

\begin{abstract} 

We report on a survey for strong (rest equivalent width $W_r \geq 1$\AA),
intervening \ion{Mg}{2} systems along the sightlines to
long-duration gamma-ray bursts (GRBs).  The GRB spectra which
comprise the survey have a heterogeneous mix of resolution and
wavelength coverage, but we implement a strict, uniform set of search
criteria to derive a well-defined statistical sample.  
We identify \nmg\ strong \ion{Mg}{2} absorbers along \ngrb\ 
GRB sightlines (nearly every sightline exhibits at least one absorber)
with spectra covering a total pathlength
$\Delta z = \pth$ at a mean redshift $\bar z = 1.1$.  In contrast,
the predicted incidence of such absorber systems along the 
same path length to quasar sightlines is only \nqmg. 
The roughly four times higher
incidence along GRB sightlines is inconsistent with a statistical
fluctuation at greater than 99.9\% c.l.  Several effects could
explain the result: 
(i) dust within the \ion{Mg}{2} absorbers
obscures faint quasars giving a lower observed incidence along
quasar sightlines; 
(ii) the gas is intrinsic to the GRB event;
(iii) the GRB are gravitationally lensed by these absorbers.
We present strong arguments against the first two effects 
and also consider lensing to be an unlikely explanation.
The results suggest that at least one of our fundamental
assumptions underpinning extragalactic absorption line research
is flawed.


\end{abstract}
]

\altaffiltext{\lick}{UCO/Lick Observatory; University of California, Santa Cruz;
	Santa Cruz, CA 95064; xavier@ucolick.org}
\altaffiltext{\uc}{Department of Astronomy; University of Chicago;
	5640 S. Ellis Ave., Chicago, IL 60637; hchen@oddjob.uchicago.edu}
 \altaffiltext{\berk}{Department of Astronomy, 601 Campbell Hall, 
        University of California, Berkeley, CA 94720-3411}
 \altaffiltext{\geneve}{Observatoire de Gen\`eve, 51 Ch. des Maillettes, 
	1290 Sauverny, Switzerland}
 \altaffiltext{\chile}{Departamento de Astronom\'ia, 
Universidad de Chile, Casilla 36-D, Santiago, Chile}
 \altaffiltext{\ioa}{Institute of Astronomy, University of Cambridge,
	Madingley Road, Cambridge, CB3 0HA, United Kingdom}
 \altaffiltext{\harvard}{Harvard-Smithsonian Center for Astrophysics, 
      60 Garden St., Cambridge, MA 02138}

\section{Introduction}

Shortly after the discovery of quasars \citep{schmidt63}, researchers
realized that one could study distant gas in the universe by
analyzing absorption lines in the spectra of these distant objects
\citep[e.g.][]{bs65}.  
Although debate persisted for many years as to whether the
observed gas was intrinsic to the quasar or at cosmological distance,
the latter view is now almost universally accepted and current research
focuses on studying
the dark matter power spectrum \citep[e.g.][]{cwb+03},
the interstellar medium of high $z$ galaxies \citep{wgp05},
metal enrichment \citep{schaye03,simcoe04}, and reionization
\citep{wbf+03}.

Upon establishing that long-duration ($t>2$s) gamma-ray bursts (GRBs)
are extragalactic \citep{mdk+97} with redshifts exceeding 
all but the most distant quasars \citep{kka+06}, researchers realized
that one could use the transient, bright afterglows to perform similar
observations as those for quasars \citep[e.g.][]{vmf03,cpb+05}.
Although the majority of analysis to date has focused on the
gas associated with the GRB host galaxy \citep[e.g.][]{mhk+02,sff03},
even the first GRB spectrum showed the presence of intervening gas
at redshifts significantly lower than the highest redshift system~\citep{mdk+97}.
The proposed applications include studying reionization at yet
greater distance than QSOs and probing the \lya\
forest on a well-behaved, power-law continuum \citep[e.g.][]{lr00,lgh+01}.

Here, we report the results from a survey of strong \ion{Mg}{2}
absorption systems.  These systems were among the first intervening absorption
lines discovered in quasar spectra because 
(i) the large rest wavelengths of the doublet 
allows for its detection in optical spectra for redshifts
as small as 0.15; and 
(ii) the doublet has a large oscillator strength and is resolved
with even low-resolution (FWHM~$\approx 5$\AA) spectroscopy.
As such, the \ion{Mg}{2} absorbers were one of the first classes
of quasar absorption line systems to be surveyed \citep{steidel92}.
Follow-up observations have shown that these absorbers trace
relatively bright galaxies \citep{lzt93,mzn+05,zib+05} and 
reside in dark matter halos 
with $M \approx 10^{12} \msol$ \citep{bmp04,php+06}.

In many of the GRB spectra acquired to date, the authors have reported
the presence of a \ion{Mg}{2} absorber with 
rest equivalent width $W_r > 1$\AA.  \cite{jhf+04} noted that the galaxies
identified with these absorbers may consistently
occur at small impact parameter ($\rho \approx 10$kpc) from the GRB sightline. 
%These authors also argued, for the GRB~030429 sightline, that the
%optical afterglow was not affected by strong gravitational lensing
%by the foreground galaxy.  
Over the past year, 
our collaboration (GRAASP\footnote{Gamma-Ray Burst Afterglows As Probes (GRAASP), http://www.graasp.org})
has obtained moderate to high-resolution observations of 
afterglows for GRB discovered by the {\it Swift} satellite
\citep{gcg+04}.  
In this Letter, we perform a search for strong ($W_r > 1$\AA)
\ion{Mg}{2} absorbers along these GRB sightlines and those reported
in the literature.
We compare the results to our recent determination of
the incidence of strong \ion{Mg}{2} systems 
along the sightlines to quasars in the Sloan Digital Sky Survey 
\citep[SDSS;][]{ppb06,php+06}.

\begin{table*}[ht]
\begin{center}
\caption{{\sc Survey Data for \ion{Mg}{2} Absorbers Along 
GRB Sightlines\label{tab:mgii}}}
\begin{tabular}{cccccccc}
\tableline
GRB &  $z_{GRB}$  & $z_{start}$ & $z_{end}$ & $z_{abs}$ & $W_r(2796$~\AA$)$ & 
$\Delta v$~(\kms) & Reference \\
\tableline
\multicolumn{8}{c}{$W_r(2796) \ge 1$~\AA\ \ion{Mg}{2} Statistical Sample} \\
\tableline
000926 & 2.038 & 0.616 & 2.0 &  &  &  & 8\\
010222 & 1.477 & 0.430 & 1.452  & 0.927  & $1.00 \pm 0.14$ & 74,000& 1\\
       & &&  & 1.156  & $2.49 \pm 0.08$ & 41,000 &  \\
011211 & 2.142 & 0.359 & 2.0 &  &  &   & 2 \\
020405 & 0.695 & 0.359 & 0.678 & 0.472 & $1.1 \pm 0.3$ & 65,000 & 11 \\
020813 & 1.255 & 0.359 & 1.232  & 1.224 & $1.67 \pm 0.02$ & 4,000  & 3 \\
021004 & 2.328 & 0.359 & 2.0   & 1.380  & $1.81 \pm 0.37$ & 97,000 & 4 \\
       & &&  & 1.602  & $1.53 \pm 0.37$ & 72,000 & \\
030226 & 1.986 & 0.359 & 1.956 & & & & 5 \\
030323 & 3.372 & 0.824 & 1.646  & & & & 7 \\
050505 & 4.275 & 1.414 & 2.0  & 1.695    & $1.98$ & 176,000  & 6 \\
050730 & 3.97 & 1.194 & 2.0  & & & \\
050820 & 2.6147 & 0.359 & 1.850  & 0.692 & $2.877 \pm 0.021$ & 192,000  & \\
       & &&  & 1.430 & $1.222 \pm 0.036$ & 113,000  & \\
050908 & 3.35 & 0.814 & 2.0  & 1.548  & $1.336 \pm 0.107$ & 147,000  & \\
051111 & 1.55 & 0.488 & 1.524  & 1.190  & $1.599 \pm 0.007$ & 45,000  & \\
060418 & 1.49 & 0.359 & 1.465  & 0.603 & $1.251 \pm 0.019$ & 124,000  & \\
       & &&  & 0.656 & $1.036 \pm 0.012$ & 116,000  & \\
       & &&  & 1.107  & $1.876 \pm 0.023$ & 50,000 &\\
\tableline
\multicolumn{8}{c}{Other \ion{Mg}{2} Systems Reported/Detected Along GRB Sightlines} \\
\tableline
970508 & 0.835 & &  & 0.767 & $0.736 \pm 0.3$ & 17,000 & 7 \\
991216 & 1.022 & &  & 0.770 & $2.0 \pm 0.8$ & 40,000 & 2\\
       & &&  & 0.803 & $3.0 \pm 0.7$ & 34,000 & \\
011211 & &&  & 0.316 & $2.625 \pm 1.418$ & 210,000 &  \\
030226 & &&  & 1.042 & $0.9 \pm 0.1$ & 109,000 & \\
       & &&  & 1.963 & $5.0 \pm 0.2$ & 2,000 &  \\
030328 & 1.522 & 0.359 & 1.497 & 1.295 & $0.42$ & 28,000 & 12 \\
030429 & 2.66 & & & 0.8418 & $3.3 \pm 0.4$ & 179,000 & 9 \\
050505 & &&  & 2.265    & $1.74$ & 134,000 & \\
050730 & &&  & 1.773 & $0.922 \pm 0.019$ & 157,000  & \\
       & &&  & 2.253 & $0.540 \pm 0.017$ & 120,000 & \\
050820 & &&  & 0.483 & $0.505 \pm 0.023$ & 213,000 & \\
050908 & &&  & 2.153  & $0.89 \pm 0.100$ & 93,000 & \\
060206 & 4.048 & 1.206 & 1.529 & 1.480  & & 179,000 & 10 \\
\tableline
\end{tabular}
\end{center}
\tablerefs{
1: \cite{mhk+02} 
2: \cite{vsf+06} 
3: \cite{bsc+03} 
4: \cite{Mir03} 
5: \cite{kgr+04} 
6: \cite{bpck+05} 
7: \cite{mdk+97} 
8: \cite{cgh+03} 
9: \cite{jhf+04} 
10: \cite{fynbo060206}
11: \cite{mpp+03}
12: \cite{mmp+06}
}
\label{tab:grb}
\end{table*}

\section{The Strong \ion{Mg}{2} Statistical Sample Along GRB Sightlines}

Owing to the transient nature of GRB afterglows, optical spectroscopy
has been obtained at many observatories with a diverse set of
instruments and instrumental configurations.  This includes our
own dataset \citep{pro_dataI06,dataII06} which is comprised
of observations acquired at the Las Campanas, Keck, Gemini, and
Lick Observatories with the HIRES \citep{vogt94}, MIKE \citep{bernstein03},
GMOS \citep{Hook04} and Kast 
spectrometers, respectively.  Nevertheless, a 1\AA\ \ion{Mg}{2}
absorber is sufficiently easy to identify that one can establish
a set of criteria that will yield a well-defined search path
and statistical sample.

The criteria imposed are: 
(i) the data must be of sufficient quality to detect both members
of the doublet at $>3\sigma$ significance;
(ii) the spectral resolution must resolve the doublet (we
demand FWHM~$< 500 \mkms$);
(iii) the search is limited to outside the \lya\ forest.
To provide a uniform comparison with low-resolution surveys,
we group all individual
\ion{Mg}{2} components within $500 \mkms$ of one another into a single
system and measure the total equivalent width of the \ion{Mg}{2}~2796\AA\
transition. 
For each of our GRB spectra and those reported in the literature,
we define a starting and ending redshift to search for \ion{Mg}{2}
absorbers, $z_{start}$ and $z_{end}$.  We define $z_{start}$
as the {\it maximum} of: $1215.67 (1+z_{GRB})/2796$, 
0.359 (to match $z_{min}$ for our SDSS survey),
and $\lambda_{min}^{SNR}/2796$, where $\lambda_{min}^{SNR}$ is the
lowest wavelength in the spectrum where $\sigma(W_r) < 0.3$\AA.
Similarly, we define the ending redshift to be the {\it minimum}
of: 3000\kms\ within $z_{GRB}$, 
$\lambda_{max}^{SNR}/2803$, and 2 (to match
the highest redshift with good statistics in the SDSS survey).
We have been 
conservative in defining these quantities and in several cases
have obtained the original spectra to verify the published results.
Table~\ref{tab:mgii} presents the value for each of the GRB 
sightlines in this survey.

\begin{figure*}
\begin{center}
\includegraphics[height=6.3in,angle=90]{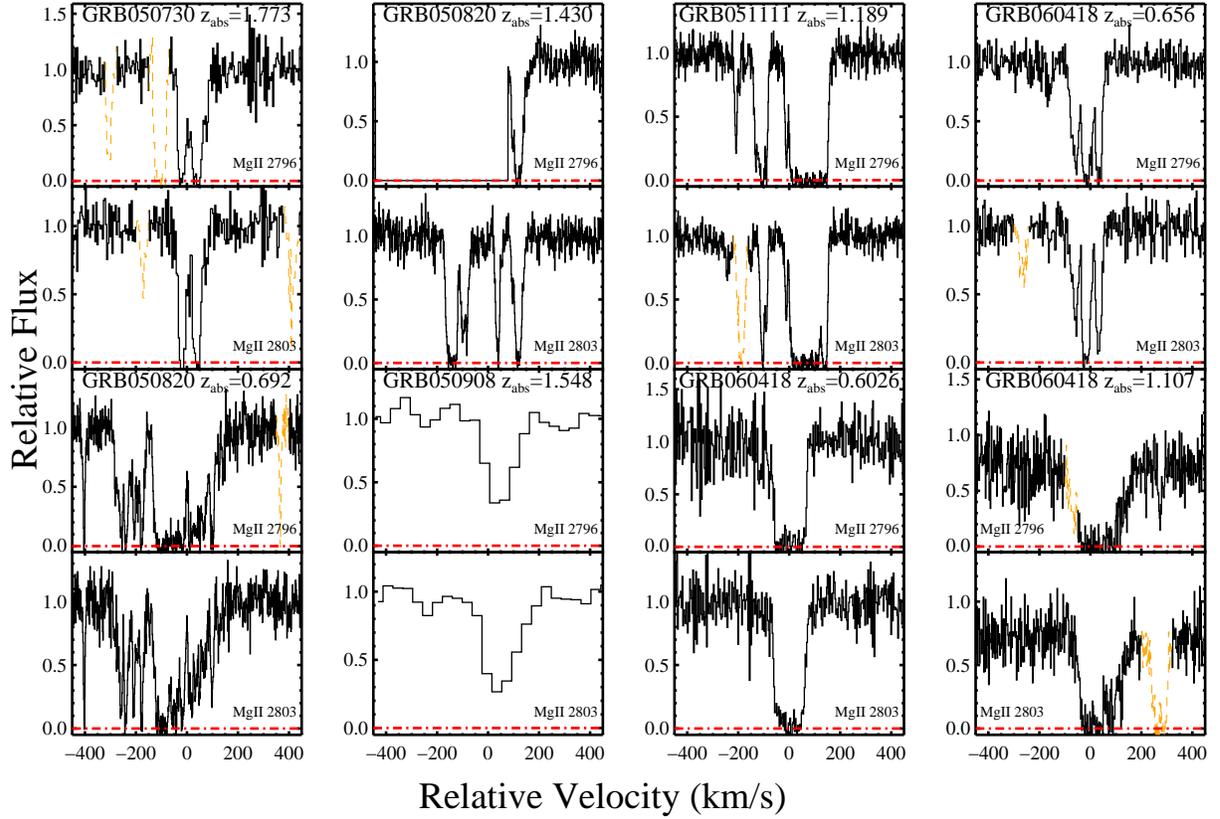}
\caption{Velocity profiles of eight of the \ion{Mg}{2} absorbers
identified along the sightlines to GRB by our GRAASP collaboration.
See \cite{pro_dataI06} and \cite{dataII06} for details of the
observations.  Dashed lines indicate features from 
coincident transitions.
}
\label{fig:gallery}
\end{center}
\end{figure*}

We have then searched these sightlines for strong \ion{Mg}{2}
absorbers and measured the rest equivalent width of the 
\ion{Mg}{2}~2796 transition.  
Figure~\ref{fig:gallery}
presents a gallery of \ion{Mg}{2} profiles from 
our GRAASP collaboration.
Details on these observations will be provided in future papers
\citep{pro_dataI06,dataII06}.  
For the literature search, we
rely on the reported equivalent width measurements. 
In nearly every case, the identification
of the \ion{Mg}{2} doublet is confirmed by the presence of strong
\ion{Fe}{2} absorption at shorter wavelengths.  
Table~\ref{tab:mgii} lists the statistical sample.  
It is astonishing that nearly every GRB sightline exhibits a
strong \ion{Mg}{2} absorber and one of those without 
shows an absorber with $W_r$ very nearly equal to 1\AA\ (GRB~050730).
Furthermore, we note that there are 
additional sightlines with insufficient spectral resolution and/or
SNR to enter the statistical sample which have very strong 
\ion{Mg}{2} absorbers ($W_r > 2$\AA).
Including these sightlines in the sample would only bolster the
results discussed below.

\section{Results and Discussion}

\begin{figure}[ht]
\begin{center}
\includegraphics[width=3.5in]{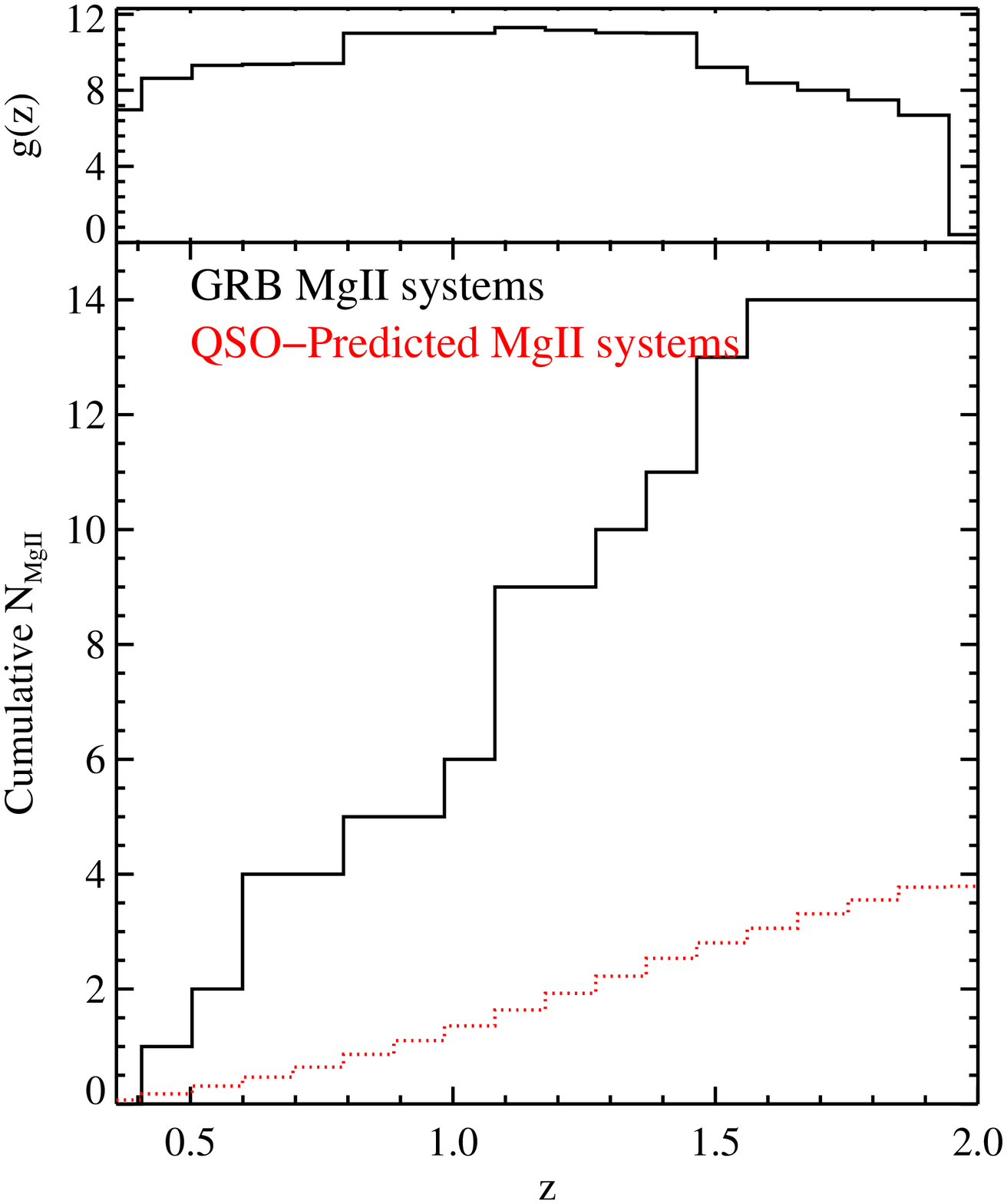}
\caption{Upper panel:  The redshift path density $g(z)$ for
the \ngrb\ sightlines which have sufficient SNR and spectral
resolution to be included in the statistical sample.  
Lower panel:  Cumulative number of \ion{Mg}{2} systems identified
along the GRB sightlines (black curve).  The red curve shows
the predicted number of systems adopting the incidence of
\ion{Mg}{2} systems $\ell^{QSO}(z)$ measured along QSO sightlines
\citep{php+06}.  The incidences observed for GRB and QSO sightlines
are inconsistent at the greater than 99.9\% level.
}
\label{fig:loz}
\end{center}
\end{figure}

In Figure~\ref{fig:loz}a we present the redshift path density
$g(z)$ which describes the number of GRB sightlines available for
a \ion{Mg}{2} search as a function of redshift.  This is a very
small sample by quasar absorption line (QAL) standards.  
In Figure~\ref{fig:loz}b, we
show the cumulative number of \ion{Mg}{2} absorbers detected
along GRB sightlines (solid line) versus the number predicted by
QSO statistics (dashed line).  
This curve was generated by convolving the $g(z)$
function for the GRB sightlines with the observed incidence
of \ion{Mg}{2} systems per unit redshift $\ell^{QSO}(z)$ from our 
survey of the SDSS \citep{ppb06,php+06}.
Our updated analysis of Data Release 4 
shows the incidence of strong \ion{Mg}{2} absorbers per unit redshift
$\ell^{QSO}(z)$ is well fitted by the following polynomial
$\ell^{QSO}(z) = -0.026 + 0.374 z - 0.145 z^2 + 0.026 z^3$
\citep{php+06}.  
Note that these results are based on over 50,000 quasars and
7,000 \ion{Mg}{2} systems with $W_r \geq 1$\AA.

\begin{figure}
\begin{center}
\includegraphics[height=3.5in,angle=90]{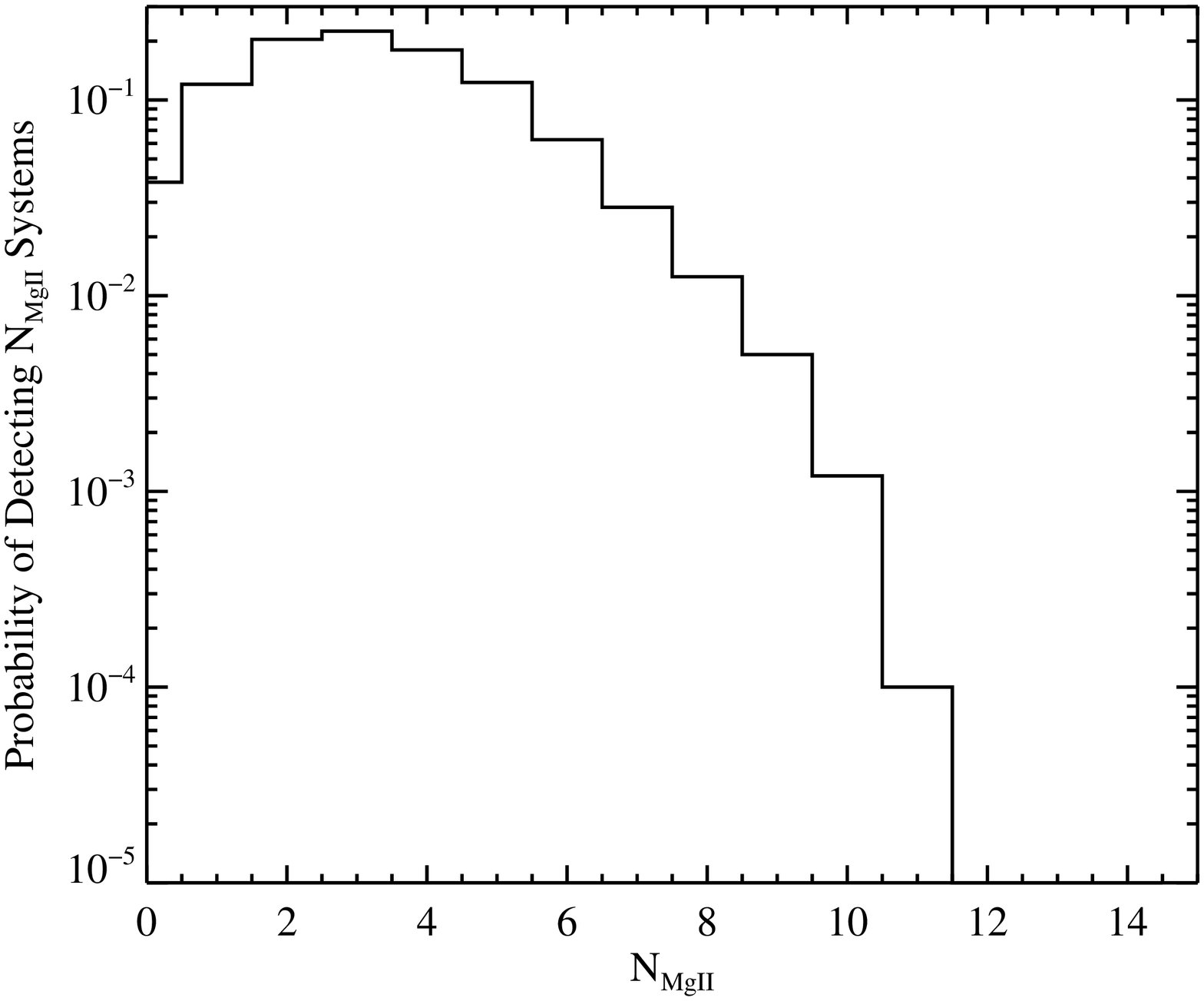}
\caption{Probability of detecting $N_{\rm MgII}$ strong \ion{Mg}{2} systems
calculated from a set of 10000 trials where we randomly drew
quasars from the SDSS dataset constrained to have nearly the
same $g(z)$ distribution as the GRB sightlines.
}
\label{fig:monte}
\end{center}
\end{figure}

An inspection of the figure 
reveals that one observes a significantly higher incidence of
strong \ion{Mg}{2} absorbers toward the GRB sightlines than along
the SDSS quasar sightlines.  Assuming Poisson statistics, the
observed incidence of \nmg\ strong \ion{Mg}{2} absorbers
is inconsistent with the average value seen towards QSOs 
at $> 99.9\%$ significance.  
We have also assessed the significance of the observation by
drawing 10000 sets of quasars from the SDSS-DR4 chosen to have 
a similar $g(z)$ function as the GRB sightlines. 
The results of this analysis is presented in Figure~\ref{fig:monte}.
We find an average of \nqmg\ strong \ion{Mg}{2} absorbers, 
that less than 0.1\% of the trials have over 10 systems, and that none
has \nmg\ absorbers.
Therefore, it seems very unlikely that
the difference in incidence between the GRB and QSO
sightlines is only a statistical fluctuation.  We note that
GRB~060418, with three strong absorption systems, is a rare object.
Monte-Carlo simulations reveal that only $2.6\%$ of randomly chosen
sets of 14 quasar lines-of-sight result in the inclusion of such a system.
Removing this GRB from consideration, however, has the combined effect of
removing both \ion{Mg}{2} systems as well as a line-of-sight, which reduces
$g(z)$, leaving little qualitative difference in the statistical result of
our analysis.

As with any astronomical survey, there are a number of associated
selection biases or possibly incorrect assumptions to the analysis.
We identify three effects which could explain the results
presented here:  
(i) dust in the \ion{Mg}{2} absorbers has obscured faint quasars
and led to a severe underestimate in $\ell^{QSO}(z)$;
(ii) the majority of the strong \ion{Mg}{2} absorbers along the
GRB sightline are not cosmological but are intrinsic to the GRB event;
(iii) GRB with bright, optical afterglows have been gravitationally
lensed by foreground galaxies hosting strong \ion{Mg}{2} absorbers.
The first effect, a selection bias, has been discussed extensively
for QAL absorbers \citep{oh84,fall93}.  
Recently, \cite{ykv+06} have shown that \ion{Mg}{2} absorbers
do impose a non-zero reddening on its quasar spectrum, but that
the average reddening for $W_r < 2$\AA\ systems is 
$E(B-V) < 0.01$\,mag.  Therefore, we consider it very unlikely
that obscuration bias is the dominant explanation.

Are the \ion{Mg}{2} absorbers along GRB sightlines
intrinsic to the GRB?  Absorption systems 
intrinsic to the quasar environment have been identified at velocities 
$\Delta v$ in excess of 50000~\kms\ \citep{jhk+96}.
These absorbers are identified because of very wide
profiles, equivalent width
variability, and/or evidence for partial covering in the doublet
line-ratios \citep[e.g.][]{bhs97}.  
Although the strong \ion{Mg}{2} absorbers show relatively wide
absorption profiles (by default) for QAL systems, the velocity widths
are less than several hundred \kms, i.e., much less than the 
implied relativistic speeds (Table~\ref{tab:mgii}).
The gas is generally highly
ionized, although there are also examples of low-ionization states.
An excellent way to test for the cosmological nature of the 
\ion{Mg}{2} absorbers along GRB sightlines is to search for the host
galaxies.  Indeed, several authors 
%\footnote{Note that 
%the GRB~030429 sightline is not included in our analysis because the
%S/N is too low and the spectral resolution too poor for the
%detection of a 1\AA\ \ion{Mg}{2} system.  We note, however, that 
%\cite{jhf+04} report a $W_r=3.3$\AA\ absorption system at
%$z=0.841$.}
have reported the identification of the host galaxies for 
strong, intervening \ion{Mg}{2} systems along GRB
sightlines \citep{mpp+03,vmf03,jhf+04}.
Only a small fraction of the strong \ion{Mg}{2} systems
listed in Table~\ref{tab:mgii} have been identified, however, and
an intrinsic origin for individual members of the sample
is not entirely ruled out.
Nevertheless, we consider it an improbable explanation at
the current time.

%For lensing:  (1) Mg2 are in massive halos; (2) Nearly every GRB
%has one of these; (3) low z GRB have lower E\_iso than higher z ones;
%(4) the impact parameter is quite small for some of the absorber
%galaxies.
%
%Against lensing:  (1) Not every deep image shows a bright (i.e. massive), 
%nearby galaxy; (2) qsos don't show much of a lensing signal;
%(3) no reports of a multiply imaged afterglow.

Are the galaxies hosting the strong \ion{Mg}{2} absorbers lensing
the background GRB events?  There are several lines of evidence
in support of this conclusion.  First, the strong \ion{Mg}{2} absorbers
reside in relatively massive dark matter halos $M \approx 10^{12} \msol$
\citep{bmp04,php+06}.  Second, the survey is biased to GRB with bright
optical afterglows, i.e.\ we are selecting a subset of the GRB population.
Third, nearly every GRB sightline
shows a $W_r > 0.5$~\AA\ \ion{Mg}{2} system.  Fourth, the impact parameter
for several of the foreground galaxies is small \citep{jhf+04}.
Fifth, the luminosities of low redshift ($z<0.5$)
GRBs appear to be significantly
lower than those of the high $z$ events.  
None of these arguments, however, is particularly strong.

Furthermore, there are a number of arguments against strong lensing.
First, estimates for the lensing
rate based on the photon number fluxes predict a small lensing
rate \citep{pm01}.  
Second, one does not always identify a bright 
foreground galaxy at small impact parameter ($<1''$) from the GRB
sightline.  We note that lensing would deflect the sightline,
perhaps by more than 10\,kpc, but that for the redshifts of interest 
here this translates to $\sim 1.3''$.    Third, it is unlikely that galaxy
or cluster lensing would provide sufficiently large magnification to 
explain the very bright afterglows.  
Finally, there have been no reports of multiple
images in late time optical follow-up observations.
For these reasons, we consider strong lensing to be an unlikely 
explanation.

\cite{frank06} have considered an alternative explanation for the 
observed effect, namely that the difference in sizes between GRBs and
QSOs leads to lower equivalent widths in QSO sightlines.  We believe, 
however, that this model is ruled out because one does not observe
unsaturated \ion{Mg}{2} lines (at high resolution) where the doublet is
not in a $2:1$ ratio \citep{churchillphd}.

In summary, we have reported on a statistically significant difference
in the incidence of strong \ion{Mg}{2} absorbers between GRB and
QSO sightlines.    Although it is partly an a-posteriori result, 
the result has the predictive test that a larger sample of GRB
sightlines will continue to show an excess of systems in comparison
with quasar sightlines.
At present, we have not identified a satisfactory
single explanation for this phenomenon.  
Our results suggest 
that at least one of our fundamental
assumptions underpinning extragalactic absorption line research
is flawed.
Before concluding, we wish to note that \cite{sr97} reported a similar
enhancement in the incidence of \ion{Mg}{2} systems along the 
sightlines to BL Lac objects, 
quasar-like phenomenon with spectra similar to GRB
that are also believed to be relativistically beamed jets. 
It may be worth considering their result in greater detail.

\acknowledgments

The authors wish to recognize and acknowledge the very significant
cultural role and reverence that the summit of Mauna Kea has always
had within the indigenous Hawaiian community.  We are most fortunate
to have the opportunity to conduct observations from this mountain.
We thank P. Madau for helpful conversations.
We acknowledge the efforts of S. Vogt, G. Marcy, J. Wright and K. Hurley
in obtaining the observations of GRB~050820.
GEP and JXP are supported by NSF grant AST-0307408. 
JXP, H-WC, and JSB are partially supported by NASA/Swift grant NNG05GF55G.

%\clearpage 

%\bibliographystyle{../../../Bibli/apj}
%\bibliography{../../../Bibli/journals_apj,../../../Bibli/grbrefs,../../../Bibli/qal,../../../Bibli/ism,../../../Bibli/instrum}

\end{document}